\documentclass[letterpaper, 10 pt, conference]{ieeeconf}  

\IEEEoverridecommandlockouts                              

\usepackage{graphicx} 
\usepackage{epsfig} 
\usepackage{mathptmx} 
\usepackage{times} 

\usepackage{interval}
\intervalconfig{soft open fences}
\usepackage{newtxmath} 
\usepackage[plainruled,
  linesnumbered,
  noline]{algorithm2e}
\usepackage{float}
\usepackage{subfigure}
\usepackage{enumerate}
\usepackage{caption}
\newtheorem{thm1}{Theorem}[section]
\newtheorem{thm2}{Theorem}[section]

\newtheorem{thm4}{Theorem}[section]

 \newtheorem{lemma}[thm1]{Lemma}
\newtheorem{definition}[thm2]{Definition}

\newtheorem{theorem}[thm4]{Theorem}

\usepackage{xcolor}

\newcommand{\R}{\mathbb{R}}

\newcommand{\buy}{b}

\newcommand{\Sel}{\mathscr{S}}
\newcommand{\Buy}{\mathscr{D}}

\DeclareMathOperator{\E}{\mathbb{E}}
\DeclareMathOperator*{\argmax}{arg\,max}
\DeclareMathOperator*{\argmin}{arg\,min}
\graphicspath{ {./figures/} }

\title{\LARGE \bf
A Nash Equilibrium Solution for Periodic Double Auctions
}

\author{Bharat Manvi$^1$ and Easwar Subramanian$^2$
\thanks{$^1$bharat.manvi@tcs.com,$^2$easwar.subramanian@tcs.com; TCS Innovation Labs, Hyderabad, India.}
\thanks{\copyright 2023 IEEE. Personal use of this material is permitted.
  Permission from IEEE must be obtained for all other uses, in any current or future
  media, including reprinting/republishing this material for advertising or promotional
  purposes, creating new collective works, for resale or redistribution to servers or
  lists, or reuse of any copyrighted component of this work in other works.}
}

\begin{document}

\maketitle
\thispagestyle{empty}
\pagestyle{empty}

\begin{abstract}
   We consider a periodic double auction (PDA) setting where buyers of the auction have multiple (but finite) opportunities to procure multiple but fixed units of a commodity. The goal of each buyer participating in such auctions is to reduce their cost of procurement by planning their purchase across multiple rounds of the PDA. Formulating such optimal bidding strategies in a multi-agent periodic double auction setting is a challenging problem as such strategies involve planning across current and future auctions. In this work, we consider one such setup wherein the composite supply curve is known to all buyers. Specifically, for the complete information setting, we model the PDA as a Markov game and derive Markov perfect Nash equilibrium (MPNE) solution to devise an optimal bidding strategy for the case when each buyer is allowed to make one bid per round of the PDA. 
   Thereafter, the efficacy of the Nash policies obtained is demonstrated with numerical experiments. 
\end{abstract}

\section{Introduction} \label{sec:intro}
\begin{figure}[H]
     \centering     
     \includegraphics[width=0.8\linewidth]{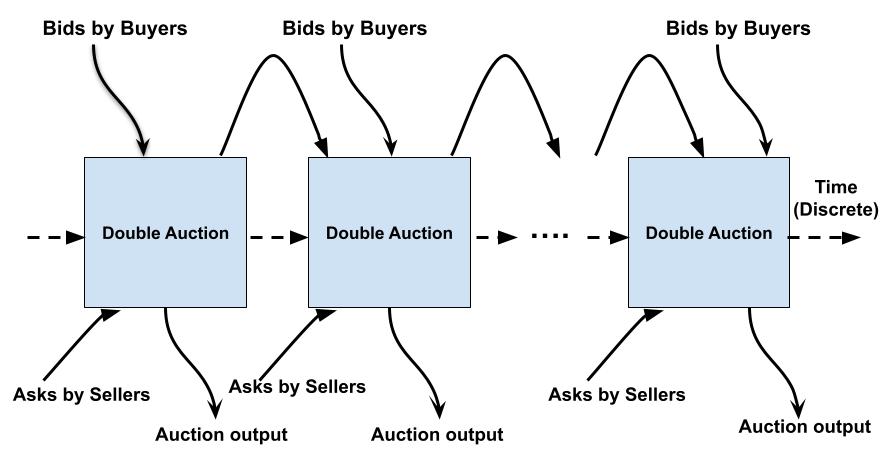}
     \caption{A Periodic Double Auction Setup}
     \label{fig:PDA}
 \end{figure}

Auctions are mechanisms that facilitate buying and selling of goods between market participants.  A double auction consists of multiple buyers and sellers submitting their asks and bids to a market institution in order to procure a target unit of a commodity. A bid or ask consists of a price-quantity pair ($p,q$) indicating that the participant is willing to buy/sell $q$ units of the commodity at a unit price $p$.  The market institution matches the buy bids with the sell asks to determine the clearing price and cleared quantities for all sellers and buyers. These type of auctions are very prevalent in stock exchanges \cite{Parsons2011} and energy markets \cite{Ketter2020}. For example, in energy markets, power generating companies are the sellers while energy brokers servicing retail customers are the buyers and a energy market regulator plays the role of the central market institution. Since the volume of trade is very high in such markets \cite{nordpool}, it is prudent to design an optimal bidding strategy on behalf of a market participant to bring in profits and system efficiency to the ecosystem. The design of such optimal bidding strategies become more pronounced in a periodic double auction (PDA) setup (see Figure \ref{fig:PDA}) wherein buyers and sellers participate in a (finite) sequence of auctions to exchange certain units of a commodity \cite{porag2018}. For example, an energy broker, armed with an estimated energy requirement for a future time slot, participates in day-ahead auctions, to procure the required energy from power generating companies by competing with other energy brokers. In these auctions, the broker will have more than one opportunity  to procure the estimated energy by participating in a sequence of auctions. For the purpose of this exposition, such an auction set up, as depicted in Figure \ref{fig:PDA}, is referred to as a periodic double auction (PDA). Evidently, in this PDA setup, an optimal bidding strategy involves planning across current and future time auctions and any small improvement in the bidding strategies of the market participants can lead to improved profits and system efficiency. Motivated by this, we formulate periodic double auctions as a Markov game and derive equilibrium solutions to devise optimal bidding strategies.

Equilibrium solutions for double auctions have been studied extensively in the past. For example, the work of Satterthwaite and Williams \cite{Satterthwaite1989} proved the existence of  non-trivial equilibria for $k$-double auctions. Analytical solutions for Nash equilibrium strategies for double auctions with average clearing price rule (ACPR) have been derived for the one buyer one seller single shot case for uniformly distributed valuations \cite{chatterjee1983} and scale based strategies \cite{susobhan20, sanjay22}. 
However, all the above approaches were developed for single shot double auctions whereas sequential decision making in a multi-agent setting was required to study PDAs. In recent years, Markov game framework have been in use  to devise bidding strategies in multi-shot auctions wherein  approaches such as multi-agent Q-learning \cite{Rashedi2016}, multi-agent  deep Q networks \cite{Ghasemi2020}, deep deterministic policy gradients \cite{Du2021} are deployed. However, much of these works do not involve deriving analytical solutions for equilibrium strategies. As far as we know, this work is the first attempt to find analytical solutions for equilibrium strategies in a PDA set up and herein lies our main contribution. 


We now elaborate on certain aspects of the  PDA considered in this work as the equilibrium analysis would depend on these specifics \cite{wilson1992strategic}.  First, we assume that the total supply available is enough to meet the overall demand requirement and that all the asks from the suppliers can be clubbed into a composite supply curve and is made known to all the buyers to make their bids. This implies that the resultant Markov game will have only the buyers as the game participants. Second, each PDA consists of fixed number of rounds, i.e., each buyer has the same fixed number of auctions to procure their respective estimated demand. Third, buyers estimate their respective procurement need  before the start of the first auction of the PDA and the estimate is not altered during the course of the PDA. Fourth, the buyers do not attempt to buy more than their outstanding requirement in any auction. Fifth, in the case that a particular buyer is not able to procure her targeted units of the commodity even after exhausting all rounds of the PDA, it will be procured outside of the auction at a higher cost.
Finally, we consider uniform payment rule (UPR), wherein, the clearing price decided through the auction mechanism is the same for all market participants. Specifically, we consider average clearing price rule (ACPR) \cite{sanjay22} where the market clearing price (MCP) is arrived as the average of the last cleared bid and last cleared ask. 

\section{Market Clearing Mechanism}\label{sec:clearprop}
 We begin by introducing a few notations.  For any positive integer $K$, let $[K]$ denote the set $\{1,\cdots,K\}$. We consider a system of $N$ buyers participating in $H$ rounds of a PDA to procure multiple units of a commodity from a set of sellers. For the purpose of this exposition, we assume that the composite supply curve from all the sellers at round $h \in [H]$ consisting of $M^h$ asks is known to all the buyers and is given by $ \mathcal{L}^{h} = \{(p^{h}_1,q^{h}_1), (p^{h}_2,q^{h}_2),\ldots,(p^{h}_{M_h},q^{h}_{M_h})\}$
where $p_h \in [0,p_{\max}]$ and $q_h \in [0,q_{\max}]$ are respectively the price and quantity components of the ask $(p^{h}_m,q^{h}_m), \textrm { with } \; m \in [M_h]$ and $p_{\max}, q_{\max}$ are suitable upper bounds for the ask.  The total supply available, at any round $h \in [H]$, is  given by, 
\begin{align}\label{eqn:total_supply}
Q^{\Sel, h} = \sum\nolimits_{m \in [M_h]} q^{h}_m .  
\end{align}

Denote the outstanding requirement of a buyer $\buy \in [N]$ at round $h$ as $Q^{{\buy},h} \ge 0$. Let $\mathcal{Q}^{h} = \{Q^{1,h},Q^{2,h},\dots,Q^{N,h}\}$ be the vector that contains the outstanding requirement of all the $N$ buyers at round $h$. Let $\mathcal{B}^h$ denote the set of all bids placed by all buyers at round $h$, wherein, each buyer $\buy \in [N]$  places at most one bid consisting of price-quantity pair $(p^{\buy,h},q^{\buy,h})$ with $p^{\buy,h} \in [0,  p_{\max}]$ and $q^{\buy,h} \in [0 , Q^{{\buy},h}]$. The number of bids placed at round $h$ is given by $B^h \le N$ and the total demand from all buyers round $h$ is given by 
\begin{equation}{\label{eqn:totaldemand}}
Q^{\Buy, h} = \sum \nolimits_{\buy \in [N]}  q^{\buy,h}     
\end{equation}
with $q^{\buy,h} = Q^{\buy,h}$. 

\begin{figure}[H]
    \centering
    \includegraphics[width=0.7\linewidth]{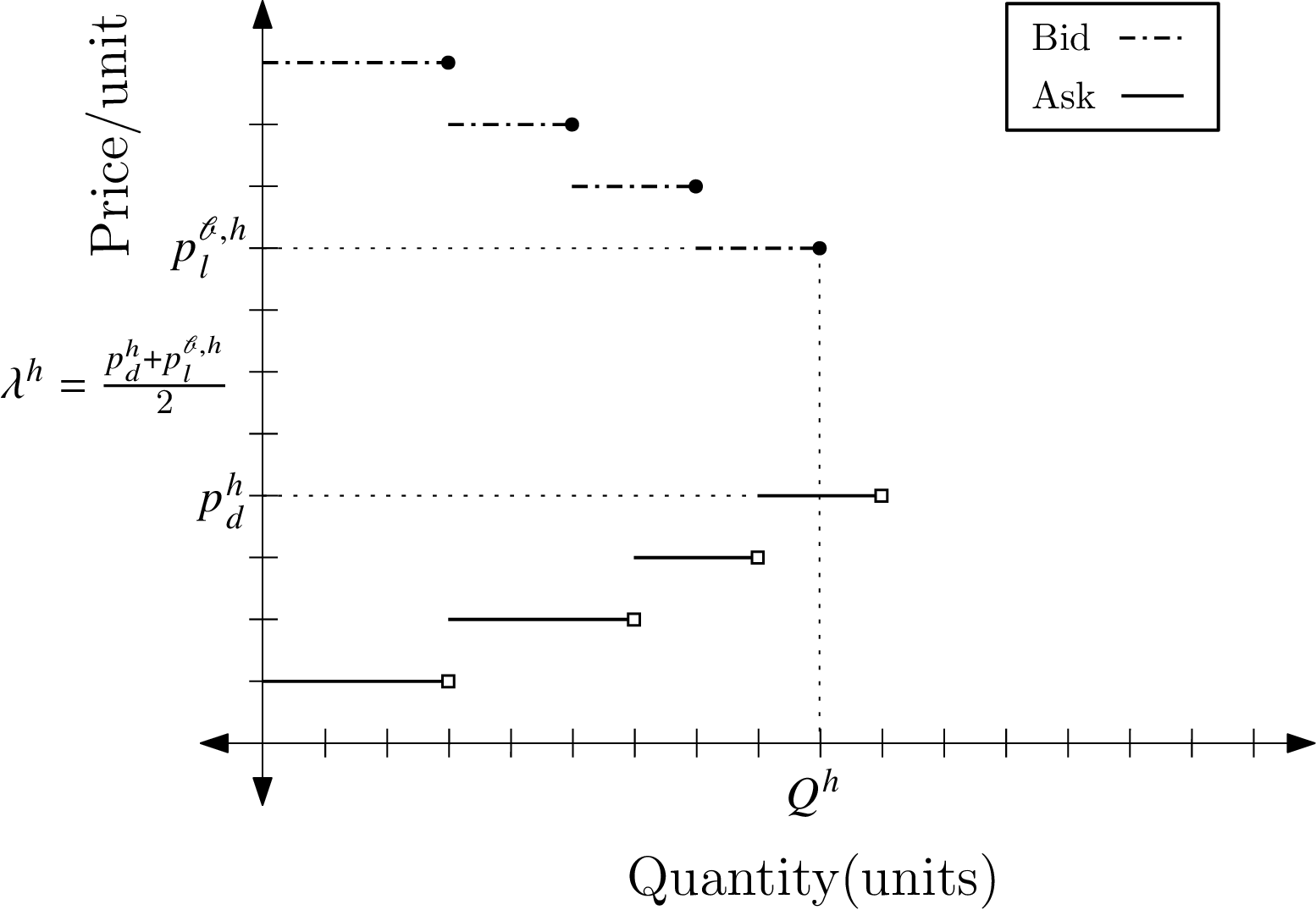}
    \caption{Average Clearing Price Rule}
    \label{fig:cleared_price}
\end{figure}

In this work, we consider average clearing price rule (ACPR) as the clearing mechanism. The ACPR is a special case of $k$-Double auction ($k \in \interval{0}{1}$), where $k=0.5$ for the ACPR. The MCP in $k$-Double auction is defined as $\lambda^h = k \cdot p^{h}_d + (k-1) \cdot p^{\buy,h}_{l}$, where $p^h_d$  and $p^{\buy,h}_l$ are the last cleared ask and bid prices respectively.   In particular for ACPR the clearing price is
$\lambda^h = \frac{p^{h}_d +  p^{\buy,h}_{l} }{2}$. Here, the bids with bid price greater than $p^{\buy,h}_l$ are fully cleared and the bid with bid price $p^{\buy,h}_l$ is either fully or partially cleared. Similarly the asks with price lesser than $p^h_d$ are fully cleared and the ask with ask price $p^h_d$ is either fully or partial cleared. The cleared quantity of the last cleared ask and bid depends on the total cleared quantity $Q^h$. Moreover, this total cleared quantity in the clearing mechanism is given as $Q^h = \min\{\sum\nolimits_{j=1}^{d} q^{h}_j, \sum\nolimits_{i=1}^{l} q^{\buy,h}_i \}$. An example of the ACPR mechanism with a cleared price and a total cleared quantity is shown in Figure \ref{fig:cleared_price}.

 \section{The Markov Game Framework}\label{sec:MGF}
Having described the clearing process of a double auction, we now model the PDA consisting of $N$ buyers with horizon $H$ as a finite horizon Markov game \footnote{A Markov game is sometimes known as a stochastic game} \cite{Zhang2023} specified by $\mathcal{M} = \langle N, S, A, C,P, H \rangle$.  The ingredients of $\mathcal{M}$ are a finite set of players $N$; a state space $S$; for each player $\buy \in [N]$, an action set $A^{\buy}$; a transition probability $P$ from $S \times A \rightarrow S$, where $A = \times_{\buy \in N } A^{\buy}$ is the action profile, with $P(s'|s,a)$ as the probability that the next state is $s' \in S$, given the current state is $s \in S$ and current action profile is $a \in A$; and a payoff function\footnote{Although, we use the term payoff, $C$ actually specifies the cost function.}  $\mathcal{C}$ from $S \times A \rightarrow \mathbb{R}^{N}$, where the $\buy$-th coordinate of $\mathcal{C}$ is $C^\buy$, is the payoff to player $\buy$ as a function of state  and action profile. 

More specifically, we let the state at round $h$ denoted by $s^h$,  to consist of $\{\mathcal{Q}^h,  \mathcal{L}^{h}\}$  and  the action $a^{\buy,h} \in A^{\buy}$ by player $\buy \in [N]$ at round $h$ consists of at most one bid belonging to the bounded set $[0,p_{\max}] \times [0,q_{\max}]$. The payoff function $C^{\buy,h}: S \times A  \rightarrow \R$ returns a scalar value to player $\buy$ specifying her cost of procurement (if any) for the auction at round $h$. More precisely, 
\begin{align*}
    C^{\buy,h}(s^h, a^h ) = \begin{cases}
          \lambda^h \cdot \alpha^{\buy,h}  & \text{at non terminal state $s^h$}\\
\Psi  \times Q^{\buy,h} & \text{when }  h=H+1,
\end{cases}
\end{align*}
where $a^h = (a^{\buy,h}, a^{-\buy,h}) \in A$ is the joint action set containing one action for each player at round $h$ with  $a^{-\buy,h}$ specifying the $N-1$ actions of all players except $\buy$. In addition, we have $\lambda^h \ge 0$  to be the clearing price of the auction at round $h$,  $\alpha^{\buy,h}$ is cleared quantity for the buyer $\buy$ at round $h$. The entity $\Psi \ge 0$ is the unit price of procuring the commodity outside of the $H$ auctions and $Q^{\buy,H+1}$ is the remaining units of the commodity to be procured by buyer $\buy$ after exhausting the $H$ rounds of the PDA. Given a state $s^h \in S$ and action profile $a^h \in A$, the next state at round $h+1$ is given by $s^{h+1} = \{\mathcal{Q}^{h+1},  \mathcal{L}^{h+1}\}$, where $\mathcal{L}^{h+1}$ refers to the uncleared asks of the supply curve from round $h$ and $\mathcal{Q}^{h+1} = \{Q^{1,h+1}, \cdots, Q^{N,h+1}\}$ with $Q^{\buy,h+1}=Q^{\buy,h}-\alpha^{\buy,h}, \forall \; \buy \in [N]$.  


At any round $h$, having seen the state $s^h$, the players choose their action based on a \textit{policy}.   A (Markov) policy for a player $\buy \in [N]$ is a collection of policies $\pi^{\buy} = \{\pi^{\buy,h} : S \rightarrow  \Delta_{A_\buy}\}_{h=1}^H$ where each $\pi^{\buy,h}(\cdot|s^h) \in \Delta_{A_\buy}$ specifies the probability of taking action $a^h \in A_{\buy}$ at state $s^h$.  
Let $\pi = (\pi^{\buy}, \pi^{-\buy})$ be the joint policy containing one policy for each player $\buy \in [N]$ where $\pi^{-\buy}$ denotes the $N-1$ policies of all players except $\buy$. The value of a joint policy $\pi$ (not necessarily Markov), at round $h$, for any player $\buy$ is a function $V^{h}_{\pi} : S \rightarrow \mathbb{R}$ defined as below.  
\begin{equation}
   V_{\pi}^h(s) = \E_{\tau \sim (P,\pi^{\buy},\pi^{-\buy})} \left[\sum\limits_{h'=h}^{H+1} C^{\buy,h'}(s^{h'}, a^{\buy,h'}, a^{-\buy,h'})  | s^h = s\right] \nonumber
\end{equation}
with $a^{\buy,h'} \sim \pi^{\buy}$, $a^{-\buy,h'} \sim \pi^{-\buy}$ and $\tau$ is a trajectory of the Markov game, generated by following the joint policy $\pi$.
As the Markov game pertaining to this work involves cost minimization as the objective, the optimal policy for any player is to find a policy that minimizes the value function. However, in a multi-agent scenario, when other players act rationally, finding optimal policy is equivalent to finding (Nash) equilibrium solution which is the best response to rational behaviour of other participating agents.  Hence, in this paper, for the PDA modelled as a Markov game, we look for MPNE \cite{abs-2011-00583,pmlr-v124-li20a} solutions defined as below. 
\begin{definition}
   Given a $N$ player finite horizon stochastic game specified by $\mathcal{M} = <N, S, A, C,P, H>$ a joint policy $\pi_* = (\pi^{\buy}_*, \pi^{\buy}_*)$ is a (Markov) perfect Nash equilibrium (MPNE) if for all $\buy \in [N]$, for all $s \in S$, for all $h \in [H]$ and for all Markov policy $\pi^\buy : S \rightarrow \Delta_{A_{\buy}}$, we have 
   \begin{equation}
 V^h_{\pi_*^{\buy}, \pi_*^{-\buy}}(s) \le V^h_{\pi^{\buy}, \pi_*^{-\buy}}(s) \nonumber
 \label{eqn:Nashcondition}
\end{equation}
\end{definition}
The perfectness of the Nash equilibrium is due to condition that the inequality in Definition \eqref{eqn:Nashcondition}  holds for every round $h \in [H]$ and for every element of the state space $S$. In the sequel, we propose a MPNE solution for the PDA problem described in Section \ref{sec:intro}. 

\section{A Nash Strategy for the Single Bid Case}\label{sec:nash}
Having elaborated on the Markov game framework, we now describe a joint policy which is an MPNE for the PDA setup considered in this work wherein each buyer is allowed to place one bid per round of the Markov game. Note that here, the goal is to find MPNE  in the space of deterministic policies. 

Recall from Equations \eqref{eqn:total_supply} and \eqref{eqn:totaldemand} that $Q^{\Buy,h}$ and $Q^{\Sel,h}$ denote total demand requirement and total supply available at round $h$. At each round $h$, let $[N^h]$ denote the set of $N$ players indexed by the decreasing order of their quantity requirement\footnote{For this work, we assume players quantity requirements are unique}. Now, let $u_h$ be the index of the ask from the set $\mathcal{L}^h$ such that all of the demand requirement at round $h$ is met. That is, 
$u_h =  \argmin\nolimits_{j}\left(  Q^{\Buy,h} \le \sum\nolimits_{m=1}^{j}q^h_m \right)$.
At round $h$, denote $Q^{\Buy_{-\buy},h} = Q^{\Buy,h} -  Q^{\buy,h}, \; \buy \in [N^h]$ as the demand requirement of all players except the player $\buy$.  Let $v^{\buy}_{h}, \; \buy \in [N^h]$ be the lowest index of the ordered set $\mathcal{L}^h$ such that the total supply available for the first $v^{\buy}_{h}$ asks satisfies the demand requirement of all players except  $\buy$. That is, $v^{\buy}_h =         \argmin\nolimits_{j}\left(  Q^{\Buy_{-\buy},h} < \sum\nolimits_{m=1}^{j}q^h_m \right) \; \forall \; \buy \in [N^h]$. Next, let us define index $z_h$ as $z_h = u_h -(H-h)$. Finally, let $\psi^h = \max\{1,\argmax_j\{v^j_h \le z_h \}\}$ as the player who bids $p_{z_h}$ and let $\phi^h$ as the player with the maximum requirement. Note that $\psi^h = \phi^h$ when $\psi^h = 1$.

The joint policy $\pi^*$ for a player $\buy \in [N^h]$ at round $h \in [H]$ for state $s^h \in S$,  can now be formulated as, 
\begin{equation}\label{eqn:nash_policy}
    \pi^{\buy,h}_*(s) = 
    \begin{cases}
        p^{\buy,h} = 0,\; q^{\buy,h} = 0    & \textrm{ if } \; Q^{\buy,h} = 0, \forall \; \buy \in [N^h] \\
     \pi_1^{\buy,h}(s)    & \textrm{ if } \; H-h \geq u_h - v^{\phi}_h \\
        \pi_2^{\buy,h}(s)  & \;  \textrm { Otherwise }
        \end{cases} 
   \end{equation}
where the policies $\pi_1^{\buy,h}(s)$ and $\pi_2^{\buy,h}(s)$ are defined as, 
\begin{align}
    \pi^{\buy,h}_1(s) = 
    \begin{cases}
        p^{\buy,h} = p_{v^{\phi}_h}  ,\; q^{\buy,h} = Q^{\buy,h} &  \textrm{ if } Q^{\buy,h} > 0, \; \buy = \phi^h  \\
    p^{\buy,h} = p_{max},\; q^{\buy,h} = Q^{\buy,h} &   \textrm{ if }  Q^{\buy,h} > 0, \; \buy \neq  \phi^h \nonumber
    \end{cases} 
   \end{align}
 \begin{align}
    \pi^{\buy,h}_2(s) = 
    \begin{cases}
        p^{\buy,h} = p_{z_h}  ,\; q^{\buy,h} = Q^{\buy,h} &  \textrm{ if } Q^{\buy,h} > 0, \; \buy = \psi^h  \\
    p^{\buy,h} = p_{max},\; q^{\buy,h} = Q^{\buy,h} &   \textrm{ if }  Q^{\buy,h} > 0, \; \buy \neq  \psi^h \nonumber
    \end{cases} 
   \end{align}  
Here, $p_{\max}$ is the maximum possible bid price and is greater than largest possible ask price \textit{i.e} $p_{\max} > p_{M_H}$. The policy in \eqref{eqn:nash_policy} suggests that the player with the highest requirement would wait for other players to get their demand satiated provided there are enough rounds as determined by $H-h \geq u_h - v^{\phi}_h$. In this case, the player with highest requirement also determines the MCP. However, if there are not enough rounds ($H-h < u_h - v^{\phi}_h$), then the player $\buy \neq \psi$ would bid for the whole quantity at the highest possible price and the player ($\psi$) would bid a price that decides the clearing price. In the case when there is only one buyer left in the market, the  policy  in \eqref{eqn:nash_policy} recommends the player to follow the supply curve. 


Having described the joint policy, we now  evaluate the value of the policy $\pi_*^{\buy,h}$ at round $h$, for player $\buy \in [N^h]$ at state $s^h \in S$. 
To this end,  the MCP $\lambda^h$, the total market cleared quantity $Q^h$ and the cleared quantity $\alpha^{\buy,h}$ for a buyer $\buy$ while adopting the policy $\pi_*^{\buy,h}$ at round $h \in [H]$ for  state $s^h$ is tabulated in the Lemma below. 
\begin{lemma}\label{lem:clearingstatistics-surplus1}
   If at round $h$, the available supply is adequate to satisfy the outstanding requirement of all players, that is, $Q^{\Buy,h} \leq Q^{\Sel,h}$ and if all the players follow the policy $\pi_*^{\buy,h}$ given as in Equation \eqref{eqn:nash_policy}, then Table \ref{tab:Cleared price and quantity} gives the clearing price and quantity for the players. 
\begingroup
\renewcommand{\arraystretch}{1.5}
\begin{table}[h]
    \centering
    \caption{Cleared price and quantities}
    \label{tab:Cleared price and quantity}
    \begin{tabular}{ |p{4cm}|p{4cm}| } 
 \hline
Case : $H-h \ge u_h - v^{\phi}_h$ & Case : $H-h < u_h - v^{\phi}_h$  \\
 \hline
 The clearing price is $\lambda_h = p_{v^{\phi}_h}$& The clearing price is $\lambda_h = p_{z_h}$\\ 
 \hline
The total market cleared quantity at round $h$ is, 
        \[Q^h_* = \min\left(\sum\limits_{j=1}^{v^{\phi}_h} q^h_j, \sum\limits_{\buy \in [N]} Q^{\buy,h}\right)\] & The total market cleared quantity at round $h$ is, 
        \[Q^h_* = \min\left(\sum\limits_{j=1}^{z_h} q^h_j, \sum\limits_{\buy \in [N]} Q^{\buy,h}\right)\]  \\ 
 \hline
 The bids placed by any player $\buy \neq \phi^h$, at round $h$ gets fully cleared. That is, 
           $\alpha^{\buy,h} =  Q^{\buy,h},\;\; \forall \buy \neq \phi$. & The bids placed by any player $\buy \neq \psi^h$, at round $h$ gets fully cleared. That is, 
           $\alpha^{\buy,h} =  Q^{\buy,h},\;\; \forall \buy \neq \psi^h$. \\ 
\hline
The bids placed the player $\buy = \phi^h$ at round $h$, gets cleared as,  \[       \alpha^{\buy,h} =  \left(Q^h - \sum\limits_{\buy \in [N] \setminus \phi^h} q^{\buy,h}_j \right)\] & The bids placed the player $\buy = \psi^h$ at round $h$, gets cleared as,  \[       \alpha^{\buy,h} =  \left(Q^h - \sum\limits_{\buy \in [N] \setminus \psi^h} q^{\buy,h}_j \right)\]\\
\hline
\end{tabular}
\end{table}
\endgroup 
\end{lemma}
\begin{proof}
First note that  policy $\pi_*^{\buy,h}$ has just two price bids with the highest bid price at $p_{\max} \geq p^h_{M_h}$. This implies that there exists at least one bid that is greater than some ask and hence the total cleared quantity $Q^h > 0$. In the case, that at round $h$, there is adequate supply to cater to the demand of all buyers, that is, $Q^{\Buy,h} \leq Q^{\Sel,h}$, the player $\phi^h$ has maximum requirement and the bid at price $p_{v^{\phi}_h}$. By construction,  $p_{v^{\phi}_h}$ is also the point where the supply and demand curve intersect and hence the MCP is $p_{v^{\phi}_h}$. It is now easy to see that, the total market cleared quantity is given by, 
\[Q^h = \min\left(\sum\nolimits_{j=1}^{v^{\phi}_h} q^h_j, \sum\nolimits_{\buy \in [N^h]} Q^{\buy,h}\right).\]
As the bids placed at the higher price $p_{\max}$ gets cleared first and since the available supply is enough to cater to outstanding demand requirement at round $h$, bids gets cleared exactly as stated in the first column  of the table in Lemma.  In similar lines, we can show for the case $H-h < u_h - v^{\phi}_h$.
\end{proof}

Having described the clearing implications for a buyer $\buy \in [N^h]$ for following the policy $\pi_*^{\buy,h}$ of Equation \eqref{eqn:nash_policy} at state $s^h \in S$, we now compute the value of the equilibrium policy for a buyer $\buy \in [N^h]$ which follows from Lemma \ref{lem:clearingstatistics-surplus1}. When $H-h \ge u_h - v^{\phi}_h$, we have, 
\begin{equation}\label{eqn:valuesurplus1}
   V^h_{\pi^{\buy}_*,\pi^{-\buy}_*}(s) = 
\begin{cases}\!
 p_{v^{\phi}_h} \times Q^{\buy,h},\; \;  & \textrm{ if } \; \buy \neq \phi^h   \\
  \begin{aligned}[b]
    \bigg[ p_{v^{\phi}_h} \times  \left(Q^h - \sum\nolimits_{\buy \in [N] \setminus \phi_h} q^{\buy,h}_j \right) \\ + \sum\nolimits_{k=h+1}^{H} p_{v^{\phi}_{k}} \times Q^{k} \bigg],
  \end{aligned} & \textrm{ if } \; \buy = \phi^h .
\end{cases}
\end{equation}
On the other hand,  when $H-h < u_h - v^{\phi}_h$, we have, 
\begin{equation}\label{eqn:valuesurplus2}
   V^h_{\pi^{\buy}_*,\pi^{-\buy}_*}(s) = 
\begin{cases}\!
 p_{z_h} \times Q^{\buy,h},\; \;  & \textrm{ if } \; \buy \neq \psi^h    \\
  \begin{aligned}[b]
    \bigg[ p_{z_h} \times  \left(Q^h - \sum\nolimits_{\buy \in [N] \setminus \psi^h} q^{\buy,h}_j \right) \\ + \sum\nolimits_{k=h+1}^{H} p_{z_{k}} \times Q^{k} \bigg],
  \end{aligned} & \textrm{ if } \; \buy = \psi^h .
\end{cases}
\end{equation}

\section{Equilibrium Analysis}
In this section, for the PDA considered in this exposition,  we show that the policy in \eqref{eqn:nash_policy} is an MPNE in the space of all deterministic policies. More precisely, we need to show that, for all $\buy \in [N^h]$, for all $s \in S$, for all $h \in [H]$ and for any deterministic policy $\pi^\buy : S \rightarrow A_{\buy}$, we have 
   \begin{equation}
 V^h_{\pi_*^{\buy}, \pi_*^{-\buy}}(s) \le V^h_{\pi^{\buy}, \pi_*^{-\buy}}(s). \nonumber
\end{equation}
Denote the bid of buyer $\buy \in [N^h]$ at state $s$ and round $h$ as prescribed by the policy $\pi^{\buy,h}_*(s)$ as $(p_*^{\buy,h}, q_*^{\buy,h})$. Further, recall that each bid of a player $\buy \in [N^h]$ belong to the bounded set $[0,p_{\max}] \times [0,q_{\max}]$ and at any round $h$, the player $\buy$ does not bid more than the outstanding demand requirement $Q^{\buy,h}$, the possible deviations available for a player $\buy$ at a state $s$ and round $h$ can be tabulated as below.
\begingroup

\renewcommand{\arraystretch}{1.5}
\begin{table}[h]
\caption{Possible Deviations}
\label{eqn:deviations}
\centering
\begin{tabular}{|cc|}
\hline
\multicolumn{1}{|c|}{Higher Priced Deviations}                                               & Lower Priced Deviations                                              \\ \hline
\multicolumn{1}{|c|}{$p^{\buy,h} > p^{\buy,h}_{*}$,  $q^{\buy,h} < q^{\buy,h}_{*}$} & $p^{\buy,h} <p^{\buy,h}_{*}$,  $q^{\buy,h} < q^{\buy,h}_{*}$ \\ \hline
\multicolumn{1}{|c|}{$p^{\buy,h} > p^{\buy,h}_{*}$, $q^{\buy,h} = q^{\buy,h}_{*}$}  & $p^{\buy,h} < p^{\buy,h}_{*}$, $q^{\buy,h} = q^{\buy,h}_{*}$ \\ \hline
\multicolumn{2}{|c|}{Equal Priced Deviation}                                                                                                                       \\ \hline
\multicolumn{2}{|c|}{$p^{\buy,h} = p^{\buy,h}_{*}$,  $q^{\buy,h} < q^{\buy,h}_{*}$}                                                                        \\ \hline
\end{tabular}
\end{table}
\endgroup
Given these deviations, we now show that, at any state $s$ and at any round $h$, a player $\buy \in [N^h]$ deviating from the policy $\pi^{\buy,h}_*(s)$ (Equation \eqref{eqn:nash_policy}) in any of the ways listed above (Table \eqref{eqn:deviations}) will not incur any less expenditure than what is accounted for via the value functions in Equations \eqref{eqn:valuesurplus1} and \eqref{eqn:valuesurplus2}. To this end, we first provide results that will be used later in the analysis. The first result provides an insight into how the MCP varies across the rounds of a PDA. 
\begin{lemma}
\label{lem:MCP}
Consider a PDA with $H$ rounds with ACPR. In the case when the composite supply curve does not change across the rounds of the PDA, the MCP at rounds $h$ and $h+1$, are related as, 
\[ \lambda^{h+1} \geq \lambda^{h}.\]
\begin{proof}
    Recall once ACPR is chosen for a PDA as the clearing mechanism at every round $h \in [H]$, the MCP is  $\lambda^h = \frac{p^{h}_d +  p^{\buy,h}_{l} }{2}$, which lies in the interval $[p_d^h, p^{b,h}_l]$ where $p_d^h$ is the price of the last cleared ask and $ p^{b,h}_l$ is the price of the last cleared bid (at round $h$). The result now follows by noting that the uncleared asks of round $h$, from which the asks of round $h+1$ would be rolled out, have prices greater than or equal to $p_d^h$. 
\end{proof}
\end{lemma}

Recall that the policy in Equation \eqref{eqn:nash_policy}, suggests that $N-1$ players to bid at price $p_{\max}$ and the remaining  player to bid at a specified (lower) price. The next result states that any deviations in the bid, by a player recommended to bid at price $p_{\max}$, at any round $h \in [H]$, might reduce the procurement cost. 
\begin{lemma}
\label{lem:pmax}
Let the conditions of Lemma \ref{lem:MCP} hold with $\Psi > \beta \cdot p_{\max}$ ($\beta > 1$) and let $\omega \in [N^h]$ be a player that is prescribed by the  policy in Equation \eqref{eqn:nash_policy}  to bid at a price $p_{\max}$ to procure his outstanding demand requirement at round $h$.  If the player $\omega$ deviates from the said policy to another policy $\pi^{\omega}$ at round $h$ at state $s$, then,  
  \begin{equation}
 V^h_{\pi_*^{\omega}, \pi_*^{-\omega}}(s) \le V^h_{\pi^{\omega}, \pi_*^{-\omega}}(s). \nonumber
\end{equation}
\begin{proof}
Among the five deviations enumerated in Equation \eqref{eqn:deviations},
the deviations suggesting that the bid price greater than $p_{\max}$ are not applicable to player $\omega$ (as by design $p_{\max}$ is the maximum bid price). The other three deviations (at round $h$) either suggest that the bid of player $\omega$ has  bid price less than equal to $p_{\max}$ or  bid quantity less than equal to $Q^{\omega,h}$. In the case $q^{\omega,h} < Q^{\omega,h}$, the bid placed by $\omega$ will lose out on the priority when compared to following the policy in Equation \eqref{eqn:nash_policy}. This implies that the bid quantity $Q^{\omega,h}$ could be partially cleared at round $h$ (as opposed to $Q^{\omega,h}$ being cleared if policy \eqref{eqn:nash_policy} is followed).  Further, by Lemma \ref{lem:MCP}, the remaining requirement of $Q^{\omega,h}$ is likely to be cleared at a higher price in  future rounds, hence the overall cost incurred by player $\omega$ is greater than or equal to the cost incurred when  policy in \eqref{eqn:nash_policy} is followed. 

Now if player $\omega$ deviates in bid price but with fixed bid quantity as $q^{\omega,h} =Q^{\omega,h}$. First consider when there are enough rounds for the player $\phi^h$ ({\textit i.e} $H-h \ge u_h - v^{\phi}_h$) and he/she deviates below the price $p_{\max}$ then the priority of the player decreases. Here, with similar arguments made earlier using Lemma \ref{lem:MCP} it can be concluded that the deviation is expensive.  Next, if the player $\phi^h$ does not have enough rounds, then the policy $\pi_2$ is recommended. Here, if the player $\omega$ has requirement greater than the player bidding at $p_{z_h}$, then for the bid price $p^{\omega,h} \in [p_{z_h},p_{\max})$, the value function is unchanged for player $\omega$. However, if the player $\omega$ bids at price $p^{\omega,h} \in [p_{v^{\psi}_h},p_{z_h})$, by construction the number of remaining rounds for the player $\buy$ would be less. Hence the player has to buy non-zero quantity from balancing market at a price $\Psi$. Furthermore, if the player $\omega$ has less requirement than the player bidding at $p_{z_h}$, then with bid price $p^{\omega,h} \in (p_{z_h},p_{\max})$, the value function is unchanged for player $\omega$. And for the bid price $p^{\omega,h} = p_{z_h}$, the value function might increase due to lemma \ref{lem:MCP}. Finally, for the bid price  $p^{\omega,h} \in [p_{v^{\psi}_h},p_{z_h})$, the condition on $\Psi > \beta \cdot p_{\max}$ will lead to higher value function for the player $\omega$. 
\end{proof}
\end{lemma}

\begin{lemma}
   \label{lem:pzh}
   Let the conditions of Lemma \ref{lem:MCP} with the balancing cost $\Psi > \beta \cdot  p_{\max}$ ($\beta > 1$) and let $\omega \in [N^h]$ be a player that is prescribed by the  policy in Equation \eqref{eqn:nash_policy}  to bid at the price $p_{v^{\phi}_h}$ or $p_{z_h}$ to procure its outstanding demand requirement at round $h$. If  the player deviates to another policy $\pi$ at round $h$, instead of following the policy in Equation \eqref{eqn:nash_policy}, then, for any state $s \in S$,   
     \begin{equation}
    V^h_{\pi_*^{\omega}, \pi_*^{-\omega}(s)} \le V^h_{{\pi^, \pi_*^{-\omega}}}(s). \nonumber
   \end{equation}
   \end{lemma}
   \begin{proof}
    Here for player $\omega$, all the five deviations listed in Equation \eqref{eqn:deviations} are possible. Now consider the case of policy $\pi_1$  which has $p_{v^{\phi}_1}$ as the bid price. Here, the deviations with the bid price greater than $p_{v^{\phi}_h}$ will increase the MCP, which leads to the increased value function.  Next if the bid price is less than $p_{v^\phi_h}$, by construction the player $\phi$ is not cleared. Similarly for bidding $q^{\buy,h} < Q^{\buy,h}$, the cleared quantity at round $h$ is less than the cleared quantity of the MPNE policy. Hence in both previous cases, by Lemma \ref{lem:MCP}, the value function increases.
    
    For the policy $\pi_2$, the recommended price is $p_{z_h}$. If the player $\omega$ bids at a price more than $p_{z_h}$, then similar to earlier case the MCP at round $h$ would be greater than or equal to $p_{z_h}$ resulting in possible increase in the cost of procurement. And, if the player bids at price $p_{z_h}$ with bid quantity less than $Q^{\omega,h}$, the cleared quantity could be less than the demand procured by following policy \eqref{eqn:nash_policy} which would imply more demand needs to be satisfied in the remaining rounds. Again from Lemma \ref{lem:MCP}, this could lead to higher cost of procurement. Finally, if the bid price is $p^{\omega,h} \in [p_{v^{\psi}_h},p_{z_h})$ then by the choice of $\Psi > \beta \cdot p_{\max}$ with suitable $\beta > 1$, the value function increases. 
   \end{proof}

\begin{theorem}
\label{thm:main}
Let the conditions of Lemma \ref{lem:MCP} hold with  the balancing cost $\Psi > \beta \cdot p_{\max}$ ($\beta > 1$). If a buyer $\buy \in [N^h]$  deviates to another policy $\pi$ at round $h$, instead of following the policy in Equation \eqref{eqn:nash_policy}, then, for any state $s \in S$ and $h \in [H]$,  
  \begin{equation}
  \label{eq:value_fn_con}
 V^h_{\pi_*^{\buy}, \pi_*^{-\buy}}(s) \le V^h_{\pi^{\buy}, \pi_*^{-\buy}}(s).
\end{equation}
\end{theorem}
\begin{proof}
From Lemmas \ref{lem:MCP}, \ref{lem:pmax} and \ref{lem:pzh} the value function of the policy \eqref{eqn:nash_policy} satisfies \eqref{eq:value_fn_con}.
\end{proof}
Note that the policy in  \eqref{eqn:nash_policy} is a Markov policy since it  only depends on the present state $s$. Moreover, the inequality \eqref{eq:value_fn_con} holding for all $h \in H$ and $s \in S$ implies that the policy satisfies sub-game perfectness.

\section{Simulations}
This section considers a simple numerical setup to demonstrate the efficacy of the Nash policies described in Section \ref{sec:nash}. 
Our setup consists of three players (buyers) in the market. The players go through a PDA simulator which has $H=24$ rounds to procure the required quantity.  The quantity requirement of the four players (\textrm{P0, P1, P2}) at some round $h \le H$ is given as $\mathcal{Q}^h=(232.18,164.6,90.7)$.  The players P0 and P2 are the players with largest and smallest  requirement respectively. The players know the supply curve (ask pattern) $\mathcal{L}_h$ which has 31 asks and the total supply $Q^{\Sel} = 1502.38 > Q^{\Buy,h} = 487.48$. We consider two values of $h$, namely $h = 1$  and $h = 23$, wherein the choice $h=1$ satisfies the condition $H-h \ge u_h - v^{\phi}_h$ and the latter does not. 

\begin{figure}[H]
   \centering
   \subfigure[]{\includegraphics[width=0.45\linewidth]{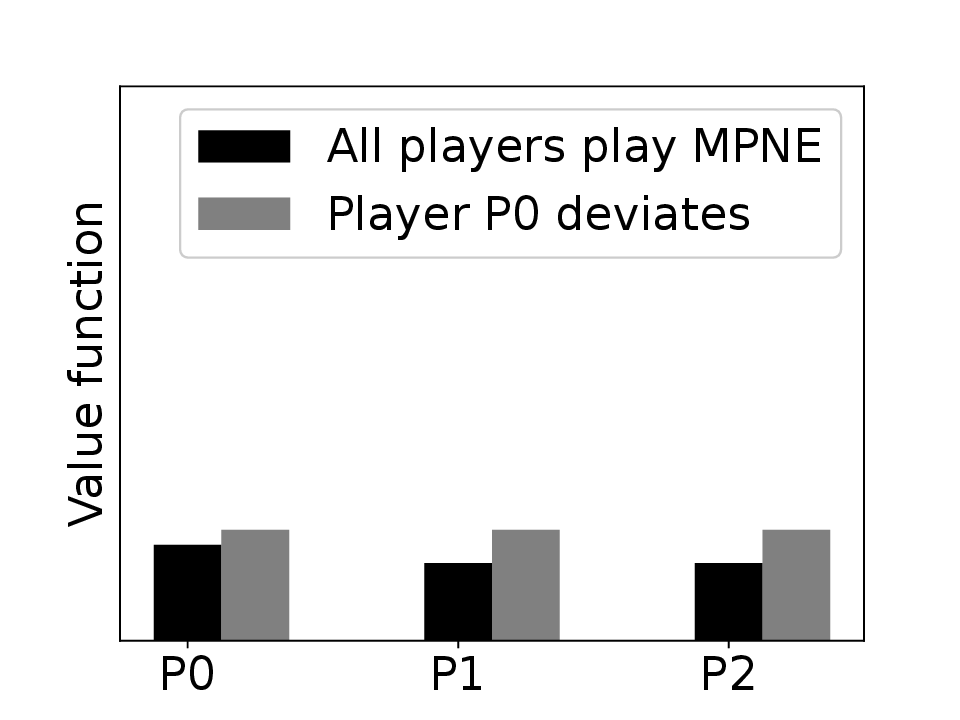}}\quad
   \subfigure[]{\includegraphics[width=0.45\linewidth]{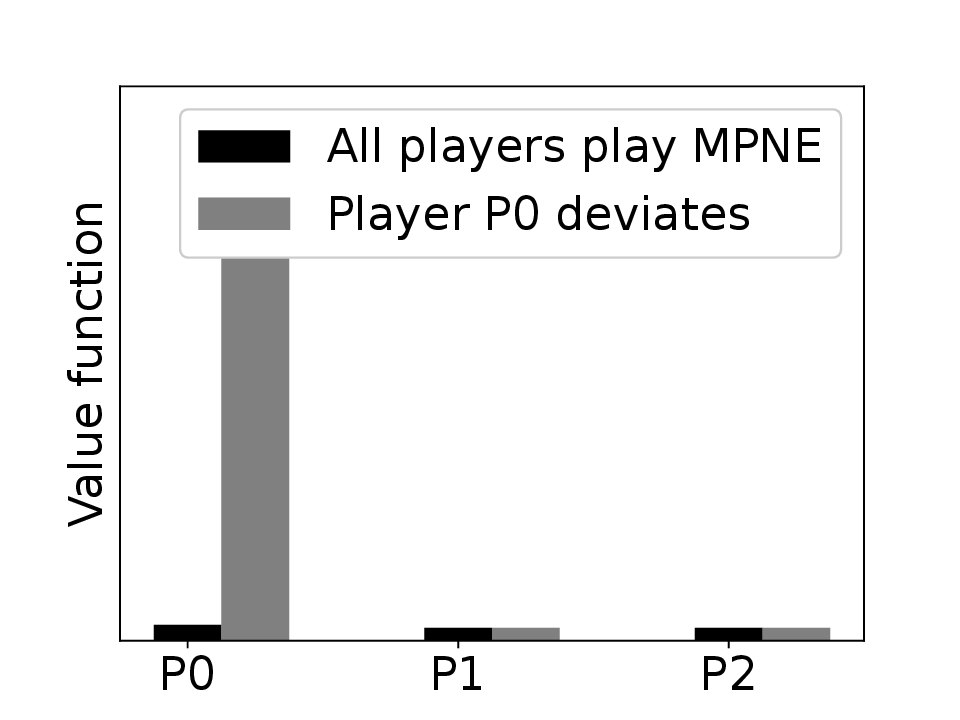}}
   \subfigure[]{\includegraphics[width=0.45\linewidth]{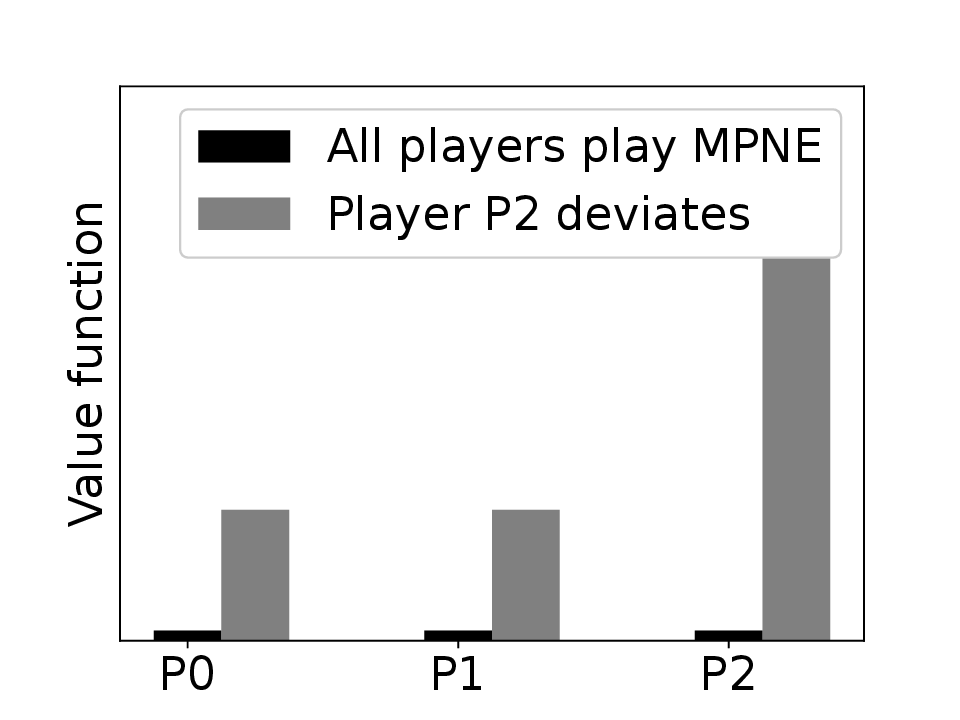}}
    \subfigure[]{\includegraphics[width=0.45\linewidth]{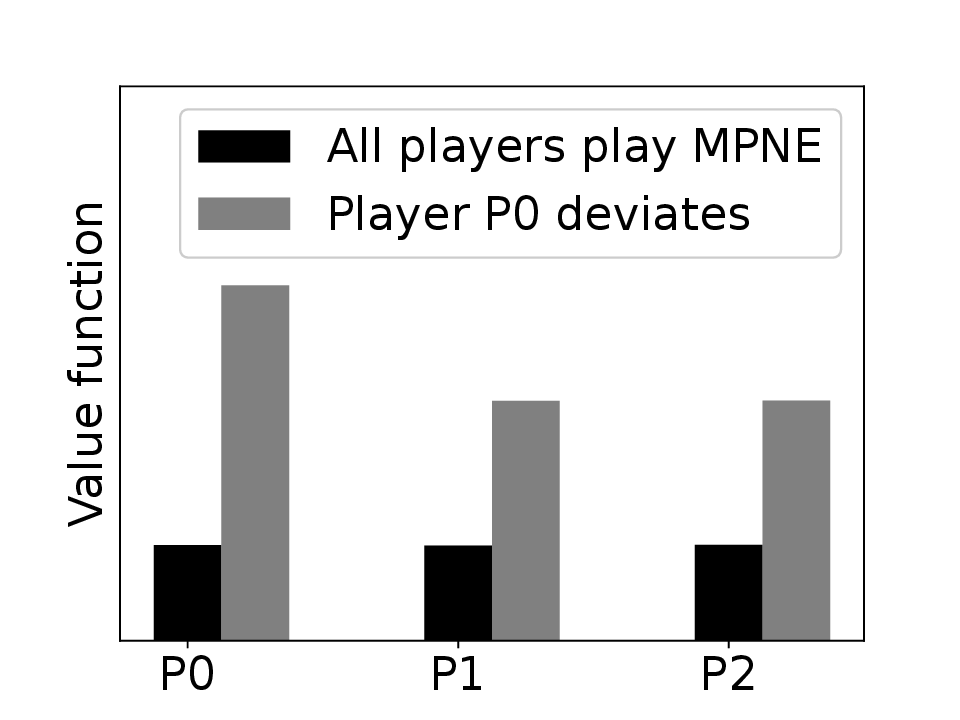}}
   \caption{Comparison of the value function of MPNE and deviation. (a) The value function for $h=1$ with the player P0 deviating to a price higher than $p_{z_h}$. (b) The value function for $h=23$ with player P0 deviating from MPNE to a lesser price than the prescribed price $p_{\max}$. (c) Value function for $h=24$ with player P2 deviating from Nash policy to bid at a price lower than $p_{z_h}$. (d) Average cost incurred by players in a series of 100 PDAs each with horizon $H = 24$ with deviation by player P0 to ZI policy}
    \label{fig:all_cases_deviation}
 \end{figure}

Figure ~\ref{fig:all_cases_deviation}(a) compares the value function when all players adopt the policy in \eqref{eqn:nash_policy} with a joint policy in which player P0 deviates to a bid price higher than the prescribed price $p_{z_h}$ at $h=1$. The higher bid price of P0 results in higher cost because of increased MCP. In Figure ~\ref{fig:all_cases_deviation}(b), for $h=23$, the condition $H-h \ge u_h - v^{\phi}_h$ is not satisfied and hence the prescribed bid price of P0 is $p_{\max}$. When P0 bids less than $p_{\max}$ its bid priority decreases resulting in procurement outside of the PDA at higher cost $\Psi$ thereby increasing the overall cost. Figure ~\ref{fig:all_cases_deviation}(c), considers the case $h=24$, wherein the condition $H-h \ge u_h - v^{\phi}_h$ is not satisfied. Here, we consider the deviation by the minimum requirement player P2 to bid at a price $p \in (p_{v^{\psi}_h},p_{z_h})$ less than the prescribed price $p_{z_h}$. This deviation to a lower price, although results in lower cost of procurement at round $h$, leads to higher overall cost as the player has to buy more units of the commodity outside of the auction at higher price $\Psi \geq \beta \cdot p_{\max}$. Finally, in Figure \ref{fig:all_cases_deviation}(d), we consider average cost incurred by the players in 100 PDAs (each with $H = 24$ rounds) with varying demand requirement. In each of these 100 PDAs, we let player P0 deviate from the prescribed Nash policy to the Zero intelligent (ZI) policy \cite{gode1993allocative} and the corresponding value functions are compared with the value function for the Nash policy. 

\section{Conclusion}
In this paper, we formulate optimal bidding strategies for a periodic double auction setting consisting of multiple buyers competing with each other to satisfy their respective demand. Each buyer has multiple opportunities to procure their need and the composite supply curve is known to all of them. The problem is modeled as a Markov game and we propose equilibrium solutions that could act as optimal bidding strategies when all buyers behave rationally. 
Apart from proving that the proposed policies are indeed MPNE, we also conducted simple numerical simulations to demonstrate the efficacy of the proposed solution framework. The PDA set up considered in this paper have applications in devising optimal bidding strategies for day-ahead electricity markets. 

Although, in this work, we have considered only the case of adequate supply with one bid per auction per buyer, we believe that the case of multiple bids per auction and inadequate supply can be handled using the techniques developed in this work.  
Despite the fact that, the equilibrium solutions proposed here are for the complete information setting, they are still important for two reasons. First, as far as we know, ours is the first work to derive analytical equilibrium solutions for multi-shot auctions. Second, these policies can be used as a baseline to compare with a policy that is obtained in an incomplete information setting, which would be a direction of our future work. 




\bibliographystyle{IEEEtran}  
\bibliography{references}  

\begin{thebibliography}{10}
\providecommand{\url}[1]{#1}
\csname url@rmstyle\endcsname
\providecommand{\newblock}{\relax}
\providecommand{\bibinfo}[2]{#2}
\providecommand\BIBentrySTDinterwordspacing{\spaceskip=0pt\relax}
\providecommand\BIBentryALTinterwordstretchfactor{4}
\providecommand\BIBentryALTinterwordspacing{\spaceskip=\fontdimen2\font plus
\BIBentryALTinterwordstretchfactor\fontdimen3\font minus
  \fontdimen4\font\relax}
\providecommand\BIBforeignlanguage[2]{{%
\expandafter\ifx\csname l@#1\endcsname\relax
\typeout{** WARNING: IEEEtran.bst: No hyphenation pattern has been}%
\typeout{** loaded for the language `#1'. Using the pattern for}%
\typeout{** the default language instead.}%
\else
\language=\csname l@#1\endcsname
\fi
#2}}

\bibitem{Parsons2011}
\BIBentryALTinterwordspacing
S.~Parsons, J.~A. Rodriguez-Aguilar, and M.~Klein, ``Auctions and {B}idding : A
  {G}uide for {C}omputer {S}cientists,'' \emph{{ACM} Computing Surveys},
  vol.~43, no.~2, pp. 1--59, Jan. 2011. [Online]. Available:
  \url{https://doi.org/10.1145/1883612.1883617}
\BIBentrySTDinterwordspacing

\bibitem{Ketter2020}
\BIBentryALTinterwordspacing
W.~Ketter, J.~Collins, and M.~de~Weerdt, ``The 2020 {P}ower {T}rading {A}gent
  {C}ompetition,'' \emph{{SSRN} Electronic Journal}, 2020. [Online]. Available:
  \url{https://doi.org/10.2139/ssrn.3564107}
\BIBentrySTDinterwordspacing

\bibitem{nordpool}
\BIBentryALTinterwordspacing
``{Nord Pool AS Anual Report},'' 2020. [Online]. Available:
  \url{www.nordpoolgroup.com/49eea7/globalassets/download-center/annual-report/annual-review-2020.pdf}
\BIBentrySTDinterwordspacing

\bibitem{porag2018}
\BIBentryALTinterwordspacing
M.~M.~P. Chowdhury, C.~Kiekintveld, S.~Tran, and W.~Yeoh, ``{Bidding in
  Periodic Double Auctions Using Heuristics and Dynamic Monte Carlo Tree
  Search},'' in \emph{Proceedings of the Twenty-Seventh International Joint
  Conference on Artificial Intelligence, {IJCAI-18}}.\hskip 1em plus 0.5em
  minus 0.4em\relax International Joint Conferences on Artificial Intelligence
  Organization, 7 2018, pp. 166--172. [Online]. Available:
  \url{https://doi.org/10.24963/ijcai.2018/23}
\BIBentrySTDinterwordspacing

\bibitem{Satterthwaite1989}
\BIBentryALTinterwordspacing
M.~A. Satterthwaite and S.~R. Williams, ``{Bilateral Trade with the Sealed Bid
  k-double Auction: Existence and efficiency},'' \emph{Journal of Economic
  Theory}, vol.~48, no.~1, pp. 107--133, June 1989. [Online]. Available:
  \url{https://doi.org/10.1016/0022-0531(89)90121-x}
\BIBentrySTDinterwordspacing

\bibitem{chatterjee1983}
K.~Chatterjee and W.~Samuelson, ``{Bargaining Under Incomplete Information},''
  in \emph{Operations Research}, vol.~31, 1983.

\bibitem{susobhan20}
\BIBentryALTinterwordspacing
S.~Ghosh, S.~Gujar, P.~Paruchuri, E.~Subramanian, and S.~Bhat, ``{Bidding in
  Smart Grid PDAs: Theory, Analysis and Strategy},'' \emph{Proceedings of the
  AAAI Conference on Artificial Intelligence}, vol.~34, no.~02, pp. 1974--1981,
  Apr. 2020. [Online]. Available:
  \url{https://ojs.aaai.org/index.php/AAAI/article/view/5568}
\BIBentrySTDinterwordspacing

\bibitem{sanjay22}
S.~Chandlekar, E.~Subramanian, S.~Bhat, P.~Paruchuri, and S.~Gujar,
  ``{Multi-Unit Double Auctions: Equilibrium Analysis and Bidding Strategy
  Using DDPG in Smart-Grids},'' in \emph{Proceedings of the 21st International
  Conference on Autonomous Agents and Multiagent Systems}, ser. AAMAS
  '22.\hskip 1em plus 0.5em minus 0.4em\relax International Foundation for
  Autonomous Agents and Multiagent Systems, 2022, p. 1569–1571.

\bibitem{Rashedi2016}
\BIBentryALTinterwordspacing
N.~Rashedi, M.~A. Tajeddini, and H.~Kebriaei, ``{Markov Game Approach for
  Multi-agent Competitive Bidding Strategies in Electricity Market},''
  \emph{{IET} Generation, Transmission and Distribution}, vol.~10, no.~15, pp.
  3756--3763, Nov. 2016. [Online]. Available:
  \url{https://doi.org/10.1049/iet-gtd.2016.0075}
\BIBentrySTDinterwordspacing

\bibitem{Ghasemi2020}
\BIBentryALTinterwordspacing
A.~Ghasemi, A.~Shojaeighadikolaei, K.~Jones, M.~Hashemi, A.~G. Bardas, and
  R.~Ahmadi, ``{A Multi-Agent Deep Reinforcement Learning Approach for a
  Distributed Energy Marketplace in Smart Grids},'' in \emph{2020 {IEEE}
  International Conference on Communications, Control, and Computing
  Technologies for Smart Grids ({SmartGridComm})}.\hskip 1em plus 0.5em minus
  0.4em\relax {IEEE}, Nov. 2020. [Online]. Available:
  \url{https://doi.org/10.1109/smartgridcomm47815.2020.9302981}
\BIBentrySTDinterwordspacing

\bibitem{Du2021}
\BIBentryALTinterwordspacing
Y.~Du, F.~Li, H.~Zandi, and Y.~Xue, ``{Approximating Nash Equilibrium in
  Day-ahead Electricity Market Bidding with Multi-agent Deep Reinforcement
  Learning},'' \emph{Journal of Modern Power Systems and Clean Energy}, vol.~9,
  no.~3, pp. 534--544, 2021. [Online]. Available:
  \url{https://doi.org/10.35833/mpce.2020.000502}
\BIBentrySTDinterwordspacing

\bibitem{wilson1992strategic}
R.~Wilson, ``Strategic analysis of auctions,'' \emph{{Handbook of Game Theory
  with Economic Applications}}, vol.~1, pp. 227--279, 1992.

\bibitem{Zhang2023}
\BIBentryALTinterwordspacing
Y.~Zhang, G.~Qu, P.~Xu, Y.~Lin, Z.~Chen, and A.~Wierman, ``Global convergence
  of localized policy iteration in networked multi-agent reinforcement
  learning,'' \emph{Proceedings of the {ACM} on Measurement and Analysis of
  Computing Systems}, vol.~7, no.~1, pp. 1--51, Feb. 2023. [Online]. Available:
  \url{https://doi.org/10.1145/3579443}
\BIBentrySTDinterwordspacing

\bibitem{abs-2011-00583}
\BIBentryALTinterwordspacing
Y.~Yang and J.~Wang, ``An overview of multi-agent reinforcement learning from
  game theoretical perspective,'' \emph{CoRR}, vol. abs/2011.00583, 2020.
  [Online]. Available: \url{https://arxiv.org/abs/2011.00583}
\BIBentrySTDinterwordspacing

\bibitem{pmlr-v124-li20a}
\BIBentryALTinterwordspacing
J.~Li, Y.~Zhou, T.~Ren, and J.~Zhu, ``Exploration analysis in finite-horizon
  turn-based stochastic games,'' in \emph{Proceedings of the 36th Conference on
  Uncertainty in Artificial Intelligence (UAI)}, ser. Proceedings of Machine
  Learning Research, J.~Peters and D.~Sontag, Eds., vol. 124.\hskip 1em plus
  0.5em minus 0.4em\relax PMLR, 03--06 Aug 2020, pp. 201--210. [Online].
  Available: \url{https://proceedings.mlr.press/v124/li20a.html}
\BIBentrySTDinterwordspacing

\bibitem{gode1993allocative}
D.~K. Gode and S.~Sunder, ``Allocative efficiency of markets with
  zero-intelligence traders: Market as a partial substitute for individual
  rationality,'' \emph{Journal of political economy}, vol. 101, no.~1, pp.
  119--137, 1993.

\end{thebibliography}

\end{document}